\documentclass[conference]{IEEEtran}
\usepackage{verbatim}
\usepackage{longtable}
\DeclareMathAlphabet{\mathpzc}{OT1}{pzc}{m}{it}
\usepackage{fancyhdr}
\usepackage{amsmath}
\usepackage{amsfonts}
\usepackage{amssymb}
\usepackage{multirow}
\usepackage[ruled,vlined,linesnumbered]{algorithm2e}
\usepackage{stfloats}
\usepackage{etoolbox}
\usepackage{float}
\usepackage{algorithmic}

\usepackage{graphicx}
\usepackage{textcomp}
\usepackage{xcolor}
\usepackage{multirow}
\usepackage{diagbox}
\usepackage{tikz}
\usepackage{enumitem}
\usepackage{url}
\usepackage{mathtools}
\usepackage{color, colortbl}
\usepackage{float}
\usepackage{subfigure}

\usepackage{cite}

\usepackage{amsmath}
\usepackage{longtable}

\usepackage{dirtytalk}
\ifCLASSOPTIONcompsoc
\usepackage[caption=false,font=normalsize,labelfont=sf,textfont=sf]{subfig}
\else
\usepackage[caption=false,font=footnotesize]{subfig}
\fi
\begin{document}
\title{A Deep Dive into the Google Cluster Workload Traces: Analyzing the Application Failure Characteristics and User Behaviors}


\author{
    Faisal Haque Bappy\\
    Syracuse University\\
    fbappy@syr.edu
    \and
    Tariqul Islam\\
    Syracuse University\\
    mtislam@syr.edu
    \and
    Tarannum Shaila Zaman\\
    SUNY Polytechnic Institute\\
    zamant@sunypoly.edu
    \and
    Raiful Hasan\\
    Kent State University\\
    rhasan7@kent.edu
    \and
    Carlos Caicedo\\
    Syracuse University\\
    ccaicedo@syr.edu
}

\maketitle

\begin{abstract}
Large-scale cloud data centers have gained popularity due to their high availability, rapid elasticity, scalability, and low cost. However, current data centers continue to have high failure rates due to the lack of proper resource utilization and early failure detection. To maximize resource efficiency and reduce failure rates in large-scale cloud data centers, it is crucial to understand the workload and failure characteristics. In this paper, we perform a deep analysis of the 2019 Google Cluster Trace Dataset, which contains 2.4TiB of workload traces from eight different clusters around the world. We explore the characteristics of failed and killed jobs in Google’s production cloud and attempt to correlate them with key attributes such as resource usage, job priority, scheduling class, job duration, and the number of task resubmissions. Our analysis reveals several important characteristics of failed jobs that contribute to job failure and hence, could be used for developing an early failure prediction system. Also, we present a novel usage analysis to identify heterogeneity in jobs and tasks submitted by users. We are able to identify specific users who control more than half of all collection events on a single cluster. We contend that these characteristics could be useful in developing an early job failure prediction system that could be utilized for dynamic rescheduling of the job scheduler and thus improving resource utilization in large-scale cloud data centers while reducing failure rates.
\end{abstract}


\IEEEpeerreviewmaketitle

\thispagestyle{fancy}
\lhead{This work has been accepted at the 10th International Conference on
Future Internet of Things and Cloud (FiCloud 2023)}
\cfoot{}

\section{Introduction}
Due to the high potential of availability, scalability, and low operation costs, large-scale data centers are quickly becoming the standard solution for a wide range of enterprise applications. But at the same time, cloud systems frequently fail due to their complexity, heterogeneity, openness, and distributed nature \cite{guan2011}. These failures can badly affect the reliability of cloud systems and also cause significant financial loss for both the user and the cloud provider. Hence, it is critical to analyze data from real-world sources in order to develop effective solutions for the challenges faced by large-scale cloud data centers.


In this paper, we conduct a deep analysis on the 2019 Google 
Cluster Trace Dataset \cite{google_2020}, which contains 2.4TiB of workload traces from 8 different clusters during the month of May 2019. This dataset has opened up new horizons for comparing the evolution of cluster managers' scheduling decisions. To explain job scheduling, the new dataset includes information on alloc sets, best-effort batch scheduler, vertical scalability, and job dependencies. In comparison to the 2011 trace dataset \cite{google_2011}, the 2019 data allows us to generalize our hypothesis across eight different clusters located across three continents and seek specific and nonconformist tendencies.

A variety of research papers have been published that address the highly granular and volatile nature of jobs  \cite{ylin2020}  \cite{liu2012}  \cite{wang2015} and their resource consumption \cite{ruan2019}  \cite{alibaba2019}. Some papers focused on the resource efficiency of cloud platforms by analyzing other publicly available cloud datasets  \cite{chen2014}  \cite{chen2018alibaba}. However, there is a scarcity of high-quality research articles focusing on creating an efficient cloud scheduler by analyzing the job-level aspects and failure characteristics. Also, for early failure detection, researchers are using different machine learning approaches to reduce unwanted resource wastage  \cite{gao2019}  \cite{pavel2017}  \cite{dabbagh2015}. But, for developing an effective and feasible failure detection system, it is important to understand the key characteristics of job failures. To the best of our knowledge, we are the first to collectively analyze 2.4TiB of data to uncover valuable insights and failure characteristics that could be useful in early failure prediction and dynamic rescheduling and thus ensuring high service reliability while minimizing failure rate and maximizing resource utilization. 

The goal of this work is to do a comprehensive analysis of the newly released cloud dataset by Google, so that, we can identify the key factors that contribute to failures and thus design an effective early failure prediction system. The next step is to perform dynamic scheduling based on the failure prediction results so that the overall performance (i.e., throughput, resource utilization, and fault tolerance) of the system can be improved. 

To accomplish our goal, we perform a four-step thorough analysis of the 2019 Google Cluster Workload Trace. The following are the major contributions of this work.

\textbf{Collection Event Analysis:} First, we analyzed the collection events to find out the impact of priority and scheduling classes in job failures.

\textbf{Failure Characterization:} We explored the nature of failed jobs and performed a failure characterization to determine the root causes. We found some significant characteristics that can be utilized to predict failures ahead of time.

\textbf{Resource Usage Analysis:} Third, we compared several key factors relevant to failed and finished jobs, such as CPU usage, memory usage, job duration, and job re-submission rate in order to better understand the significance of these characteristics.

\textbf{User Event Analysis:} To identify the heterogeneity in collection events submitted by users, we performed a novel usage analysis which unveils the usage behavior of users possessing more than 50\% of the collection events. 

The rest of the paper is organized as follows. In Section II, we present some related work on cloud data analysis and failure characterization. Then, in Section III, we provide a high-level overview of the 2019 Google Cluster Workload Trace. Section III also describes the collection events in the Google Cluster dataset and shows the failure distribution across clusters. Following that, in Section IV, we present the failure characteristics and make comparisons between failed and finished jobs. In section V, we present our analysis of user behavior. In section VI, we present some future directions in the field and finally, we conclude the paper in Section VII.

\section{Related Works}
Several prior studies have attempted to analyze the publicly available cloud dataset and find new insights about resource usage, scalability, and diverse workloads. Different researchers performed the analysis from different viewpoints. Most of them have worked with Google Cluster Dataset 2011~\cite{google_2011}. We are the first to do a deep analysis on the 2019 Google cluster dataset. 

Chen et al.~\cite{chen2014} presented some statistical properties of job and task failures by correlating them with key scheduling constraints, node failures, and attributes of users on the Google Cluster Dataset 2011. They also explored the potential for early failure prediction and anomaly detection for the jobs. Similarly, Garraghan et al.~\cite{ghan2014} used the Google Cluster Dataset 2011 to perform failure analysis and discovered that workload and server failure characteristics are highly varied, and that production cloud workloads may be accurately described using a Gamma distribution. 

Ruan et al.~\cite{ruan2019} characterized the jobs, tasks, machines, and resource usage through a comparative study and presented interesting ﬁndings about the architecture, jobs, tasks, and resource usage. Reiss et al.~\cite{reiss2012} showed that heterogeneity in Google Cluster Trace Data 2011 reduces the effectiveness of traditional slot-based and core-based scheduling and also pointed to the necessity of new resource management approaches.

Guo et al.~\cite{alibaba2019} focused on the resource efficiency of the Alibaba cloud and concluded that memory has become the new bottleneck in Alibaba's data center, limiting resource efficiency. Chen et al.~\cite{chen2018alibaba} presented a statistical proﬁle of workload pattern and clustering based on numerous common aspects of the Alibaba trace~\cite{tracealibaba2018} information. Even though all of these works attempted to characterize failures and workload patterns by studying scheduling constraints and resource consumption, they did not consider user behavior and collection events for determining workload patterns. In contrast, we considered both collection events and user behavior for understanding the actual characteristics of diverse workloads. Also, we performed our analysis on the Google Cluster Workload Traces 2019. We believe the new workload traces of the Google cloud has unveiled more vital characteristics of job failure. To the best of our knowledge, we are the first to perform a deep analysis on this dataset considering multiple perspectives. 


\section{Dataset Overview}
The Google cluster includes a set of eight clusters located in North America, Europe, and Asia. Google recorded the trace data over the entire month of May 2019. A cluster is made up of many cells, which are groups of machines connected by network switches and managed by the same cluster management system~\cite{borg2015}. Every cluster in this dataset contains approximately 12,000 machines with varying degrees of heterogeneity in their micro-architecture and chipset versions. These clusters support job assignments. Each job consists of one or more tasks, and alloc sets, which represent resource reservations for later job scheduling. To preserve confidentiality, Google researchers obfuscated personally identifiable information and transformed specific attributes to conceal system architecture information.

The overall structure of the Google cluster trace dataset can be described in four major parts: Machine Events, Instance Events, Collection Events and Resource Usage. A machine event is defined as the addition, removal, or update of a machine. Most clusters have a fairly homogeneous distribution of machine events. A collection event denotes one of eleven possible occurrences in the lifecycle of a task or alloc set. 


\subsection{Machine Lifecycle}
As machines are not homogenous, they have a wide range of microarchitectures and chipset variants with different clock rates, memory speeds, and core counts. So, it is crucial to understand how machines communicate with cluster managers in order to preserve critical characteristics like service reliability and availability. 

A typical machine lifecycle is as follows: machines become available to the scheduler to host tasks and alloc instances. While being operational, they may experience changes in resource availability. They might get removed and become unavailable due to a system failure or scheduled maintenance. Then after resolving the problem, they become available to the cluster again. 

\subsection{Collection Events}
\begin{figure}[]
  \centering
  \includegraphics[width=\linewidth]{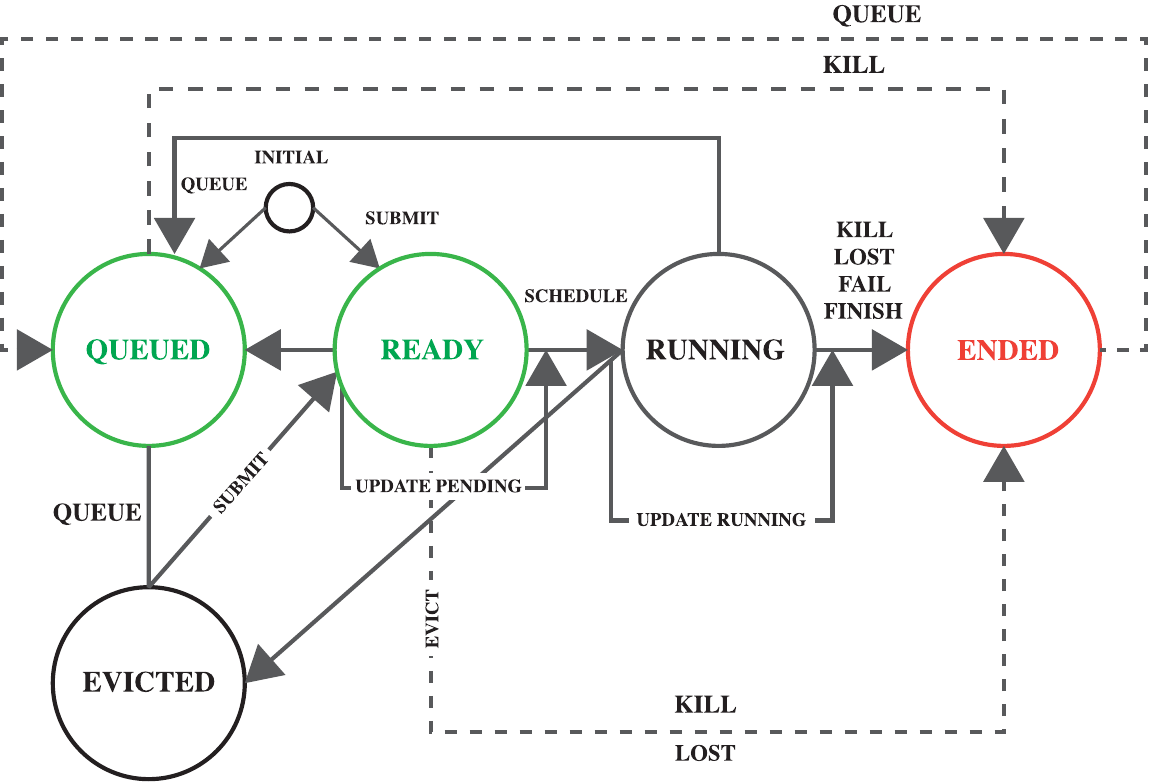}
  \caption{State Diagram of Collections}
  \label{fig:state-diag}
\end{figure}

Collections are referred to as jobs or alloc sets by Google. The collection lifecycle is made up of many state changes and various collection events~\cite{google_2020}. Figure~\ref{fig:state-diag} displays all conceivable situations for a collection's lifecycle. The dotted lines represent rare transitions, while the solid lines represent regular transitions. The green circles indicate the source state, and the red circles represent the sink state. 

\begin{figure}[]
  \centering
  \includegraphics[width=\linewidth]{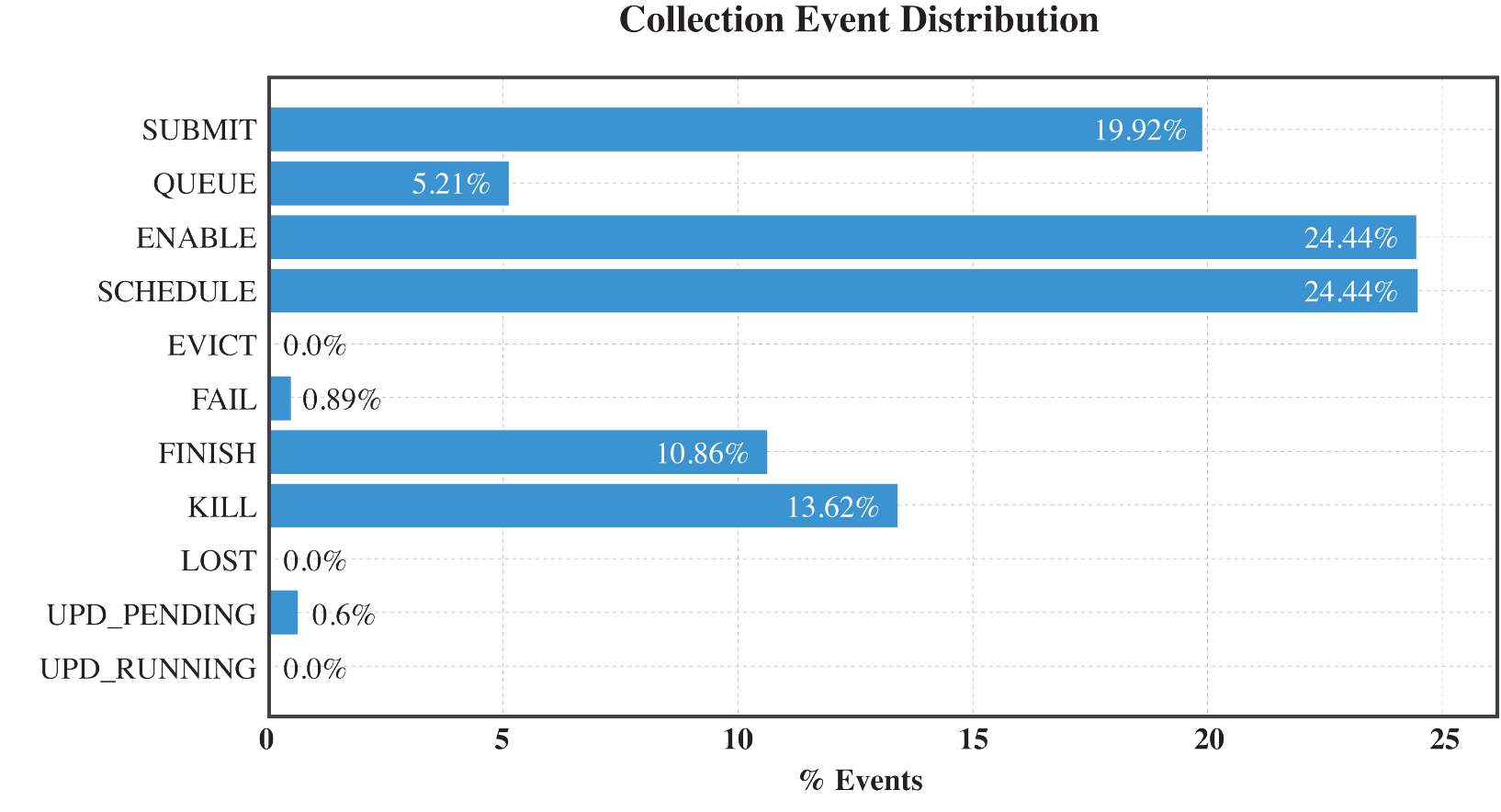}
  \caption{Collection Events Distribution}
  \label{fig:collection-dist-a}
\end{figure}

\begin{figure}[]
  \centering
  \includegraphics[width=0.7\linewidth]{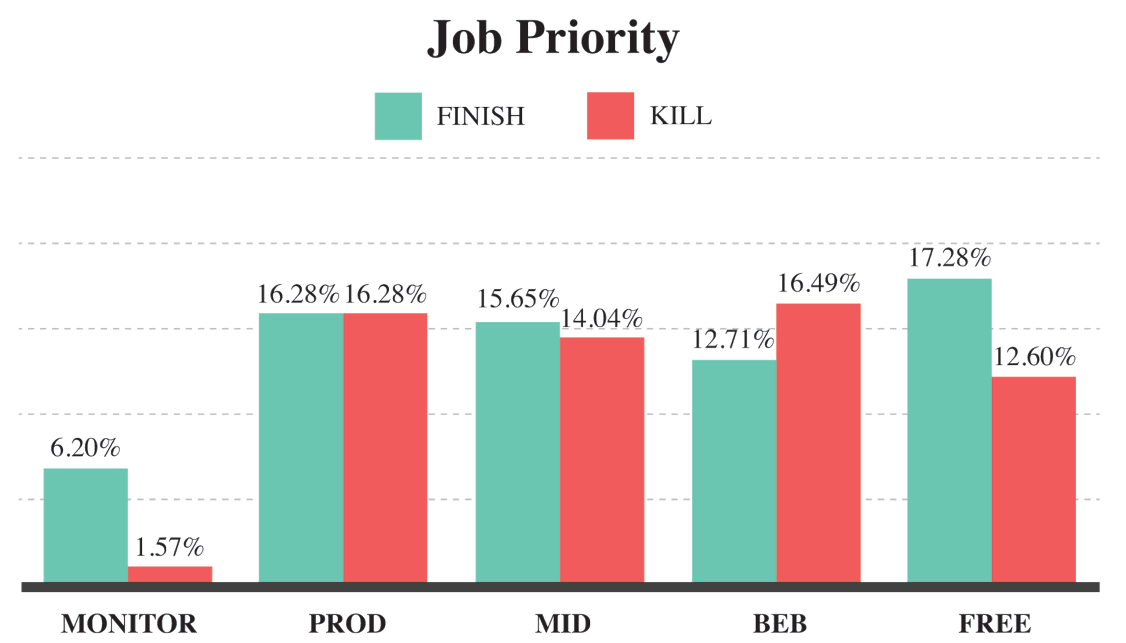}
  \caption{Job Priority distribution among Finished and Killed Jobs}
  \label{fig:collection-dist-job-priority}
\end{figure}

\begin{figure}[]
  \centering
  \includegraphics[width=0.7\linewidth]{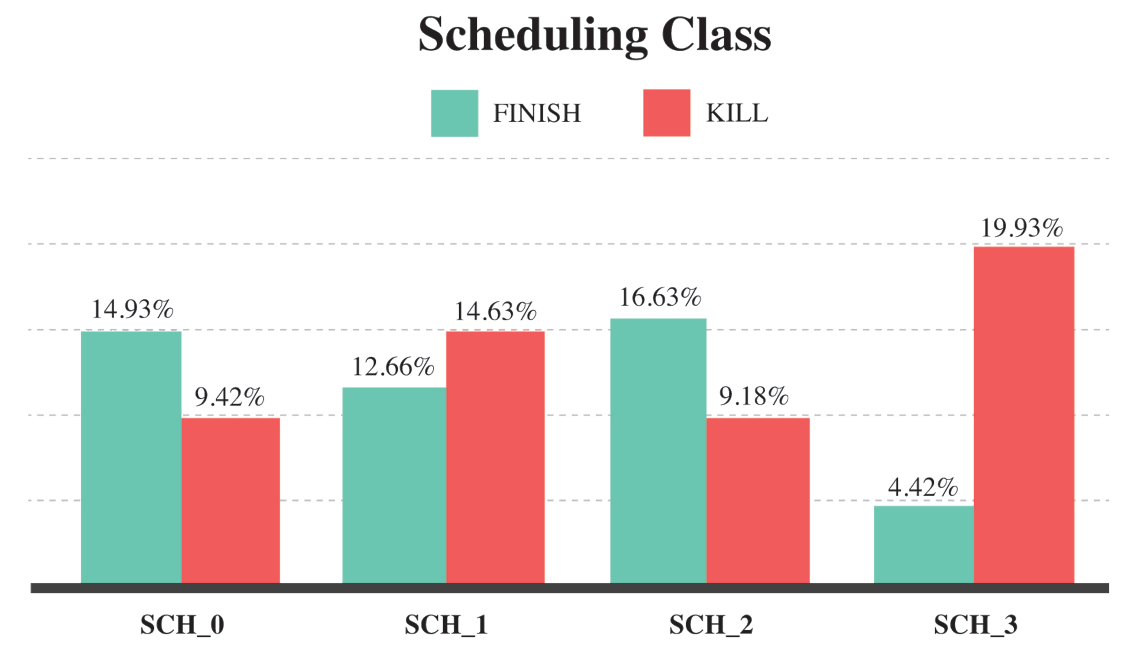}
  \caption{Scheduling Class distribution among Finished and Killed Jobs}
  \label{fig:collection-dist-sc-class}
\end{figure}

Every collection has its own set of needs for latency expectations, resource reserve, resource preference, and resource request handling. Google allows users to specify their latency expectations for job execution~\cite{borg2015}. Users could reserve resources for a later job run in addition to the scheduling class. Five distinct job priorities handle resource preference. Each work priority has its own price policy and service level goals (SLO). Users can also employ vertical scaling while conducting jobs on Google. Vertical scaling lowers spare resources efficiently by deciding how much CPU and RAM to request~\cite{borg2020}. When all of these expectations and requirements come together, ensuring proper resource allocation, job assignment, and service quality becomes extremely difficult. Figure~\ref{fig:collection-dist-a} presents the distributions of different types of jobs in all clusters. Figure~\ref{fig:collection-dist-job-priority} presents the distribution of Job Priority of finished and killed jobs. And, Figure~\ref{fig:collection-dist-sc-class} presents the distribution of Scheduling Class of finished and killed jobs. 



\begin{figure}[]
  \centering
  \includegraphics[width=\linewidth]{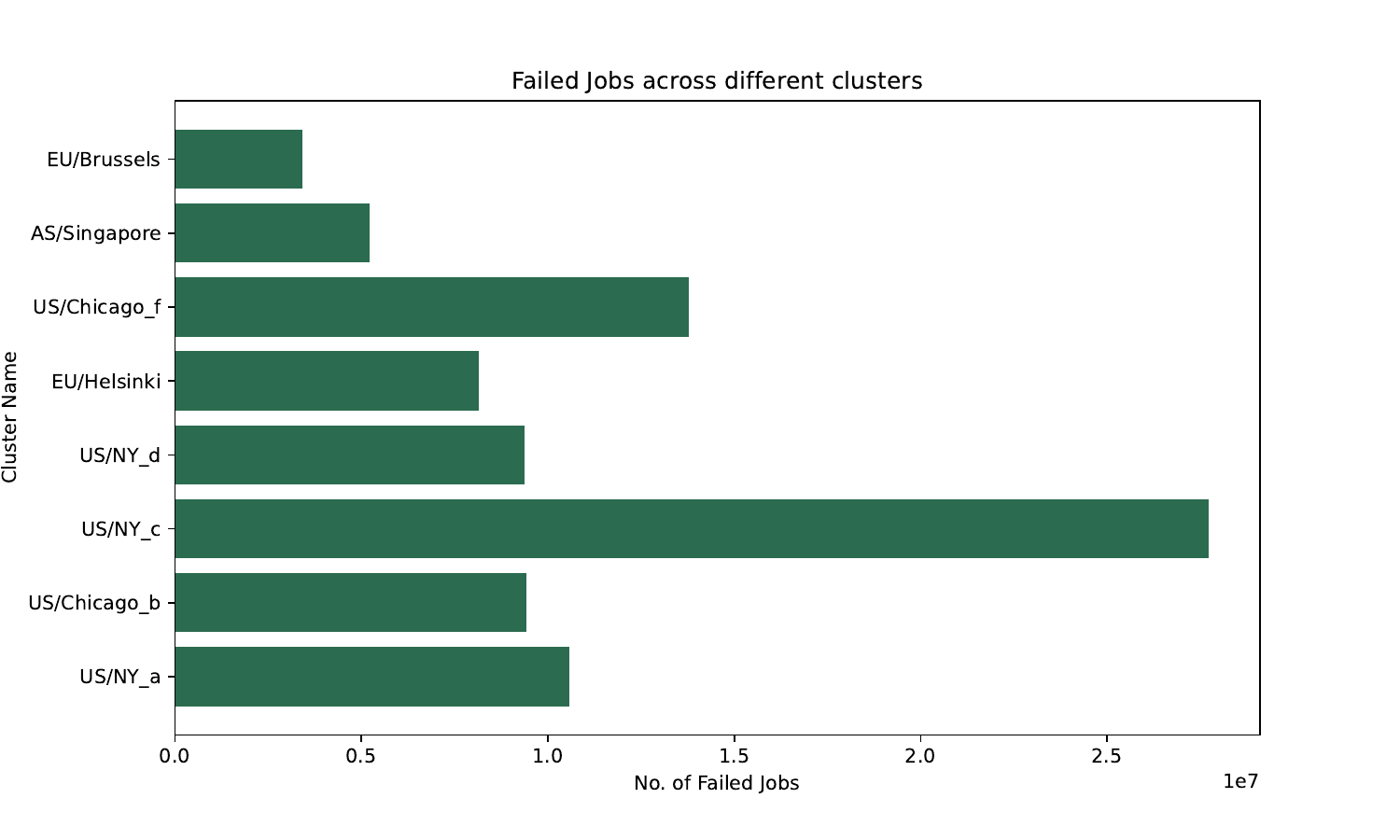}
  \caption{Failed Jobs across different Clusters}
  \label{fig:failed-jobs-dist-a}
\end{figure}

\section{Failure Characterization}
In this section, we characterize the major causes of job failures. We can infer from the distributions of failed and killed jobs (Figure~\ref{fig:collection-dist-a}) that the clusters are encountering a significant number of job failures on regular basis. These failures hinder resource utilization and have a negative impact on the cloud platform's efficiency. Hence, it is essential to understand and address the characteristics of common failures.

\subsection{Resource Allocation}
When a job is submitted, the associated cluster allocates a fixed number of resources for the job (in this case CPU and Memory). We observe that there are several cases where the allocated resource is not enough for running the job. In some cases, the job requires two or three times more resources than the allocated resources. 

\subsubsection{CPU}
\begin{figure}[]
  \centering
  \includegraphics[width=3.4in]{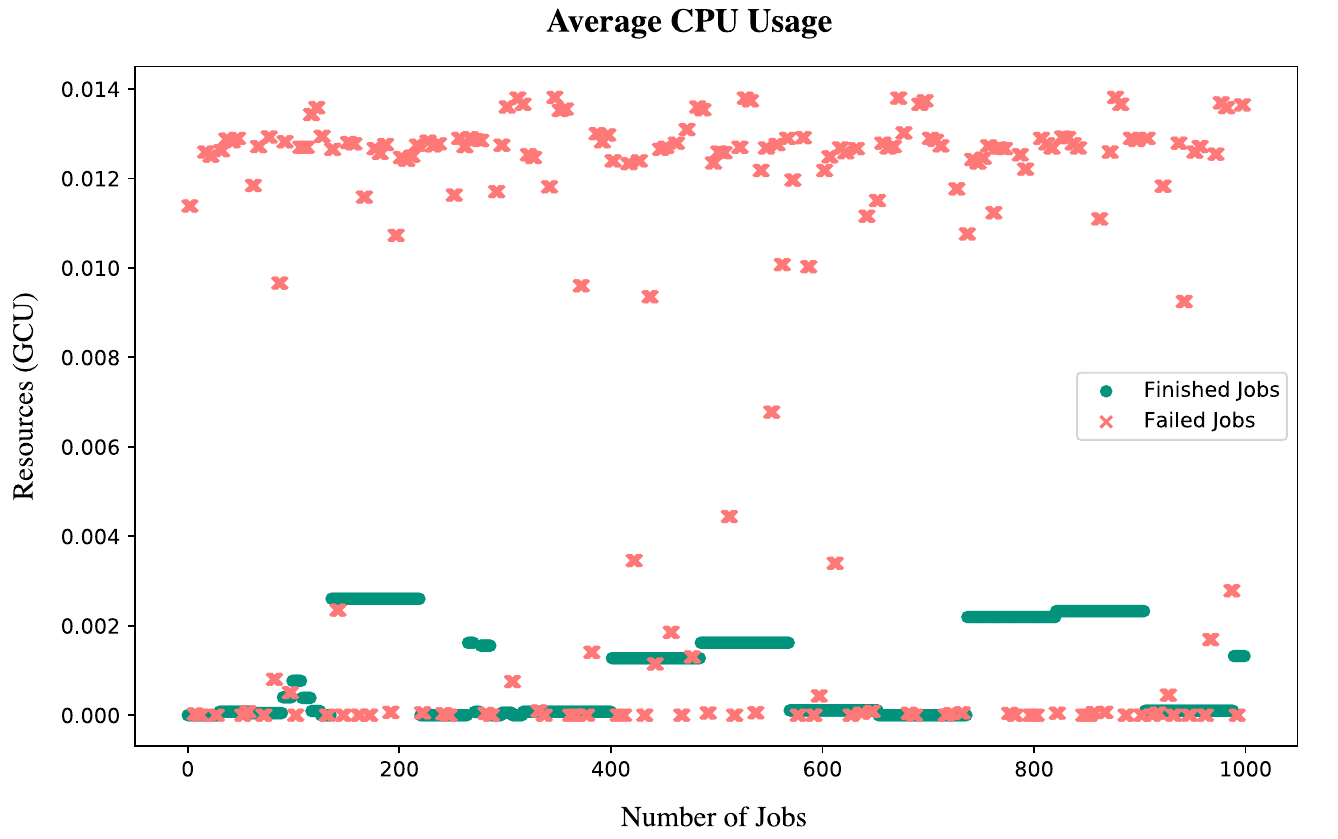}
  \caption{Average CPU Usage Comparison}
  \label{fig:cpu-1}
\end{figure}

For understanding the usage behavior, we compared the average CPU usage of failed and finished jobs. Figure~\ref{fig:cpu-1} shows that most of the failed jobs use more CPU resources than finished jobs. In some cases, the usage is more than 13 times higher. So, this common pattern of higher CPU usage is an indicator of failed jobs. This can be utilized to detect potential failures and reduce the wastage of resources. 

\begin{figure}[]
  \centering
  \includegraphics[width=\linewidth]{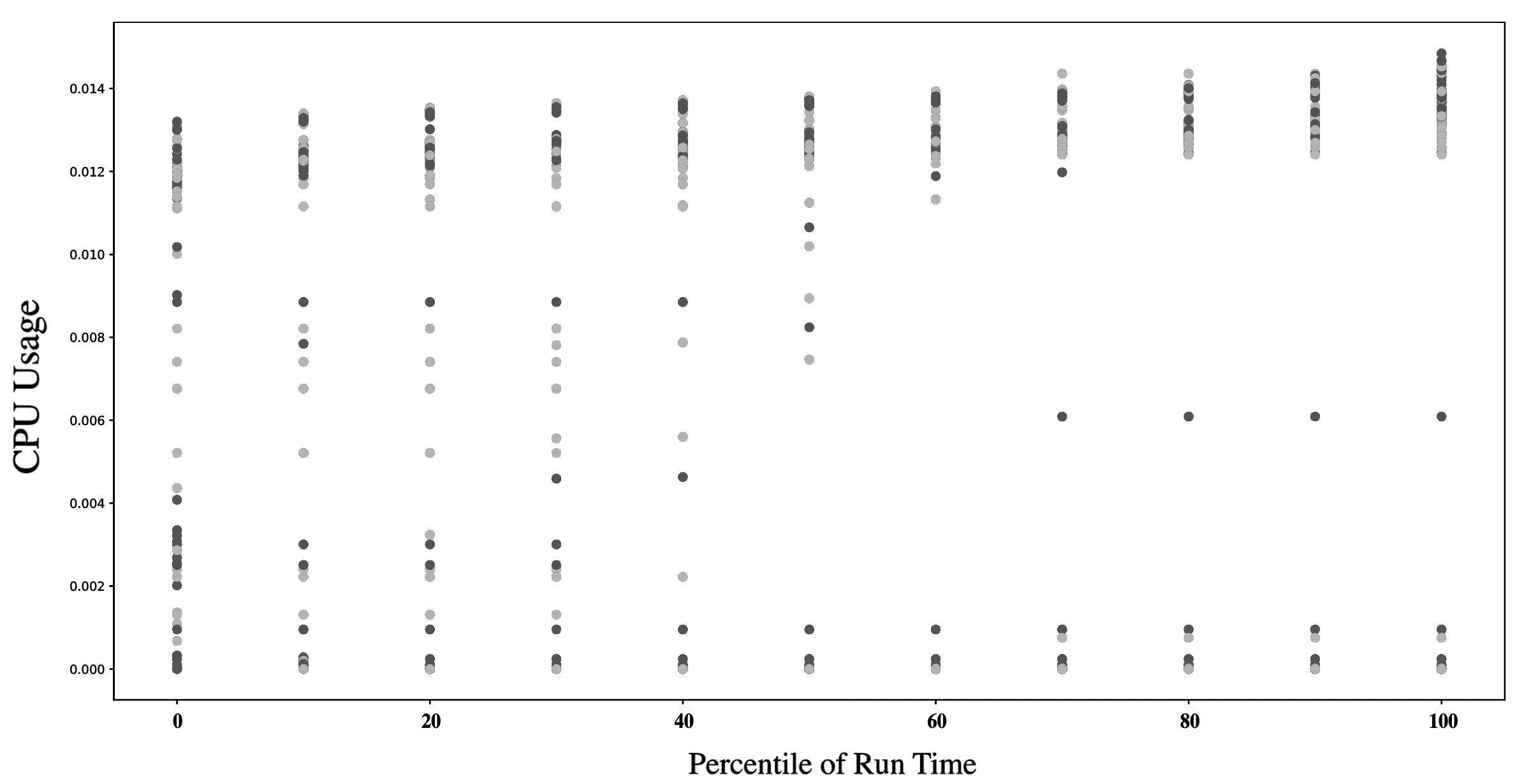}
  \caption{CPU Usage Distribution}
  \label{fig:cpu-2}
\end{figure}

We also analyze the CPU usage on different percentiles of their run-time and found that the user becomes higher and higher with the percentile of run-time. Specifically, the usage after the $60^{th}$ percentile is much higher than on any previous occasions (Figure \ref{fig:cpu-2}). At the beginning of a job, the usage seems average as other jobs. But as time passes the usage becomes significantly high for failed jobs. 

\subsubsection{Memory}
\begin{figure}[]
  \centering
  \includegraphics[width=3.4in]{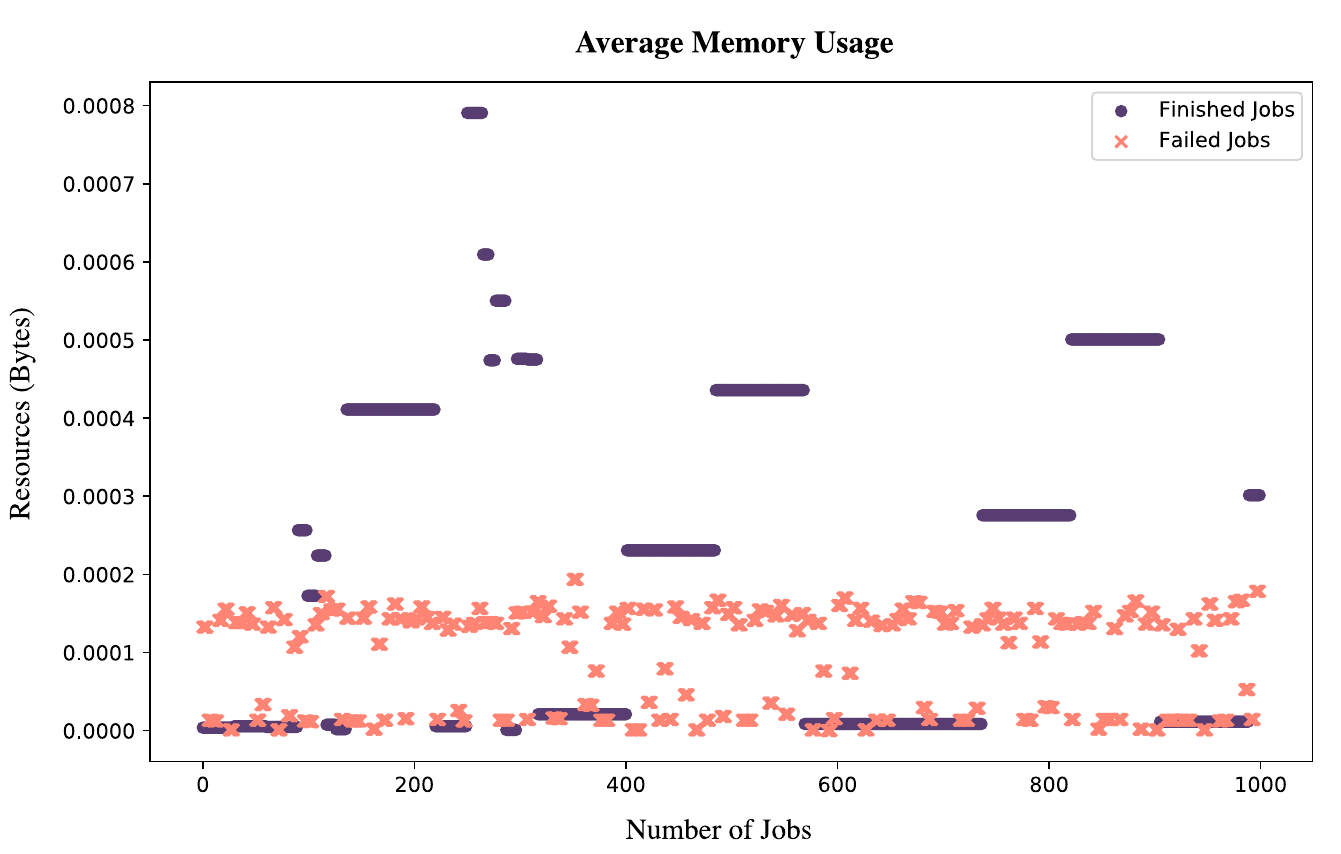}
  \caption{Average Memory Usage Comparison}
  \label{fig:memory}
\end{figure}

In the case of memory usage, the situation is a bit different. Usually, the finished jobs have a diverse range of memory usage. But in failed jobs, there is a common pattern between memory usage. Figure~\ref{fig:memory} shows that in most cases failed jobs are consuming less memory than finished jobs and there is less diversity in their usage. Though most of the failed jobs require more resources, in the execution period they only consumed near or below the average usage. 

\begin{figure}[]
  \centering
  \includegraphics[width=3.4in]{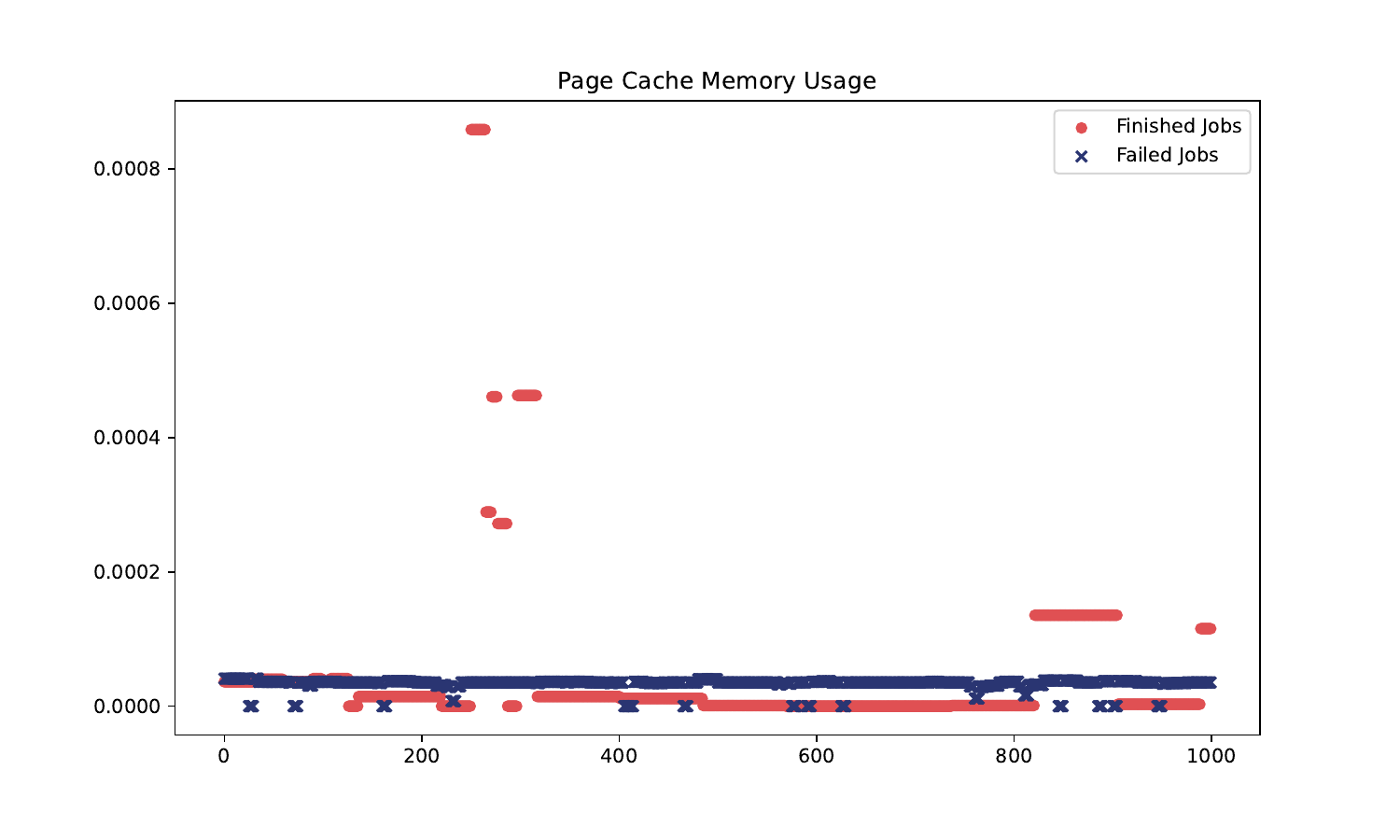}
  \caption{Page Cache Memory Usage Comparison}
  \label{fig:pc-memory}
\end{figure}

We also analyzed the Page Cache Memory usage of failed and finished jobs. Figure \ref{fig:pc-memory} shows the comparison of page cache memory usage between finished and failed jobs. And there is no significant memory consumption from failed jobs. That means most of them are failing before using cache memory. 

\subsection{Job Duration}
\begin{figure}[]
  \centering
  \includegraphics[width=\linewidth]{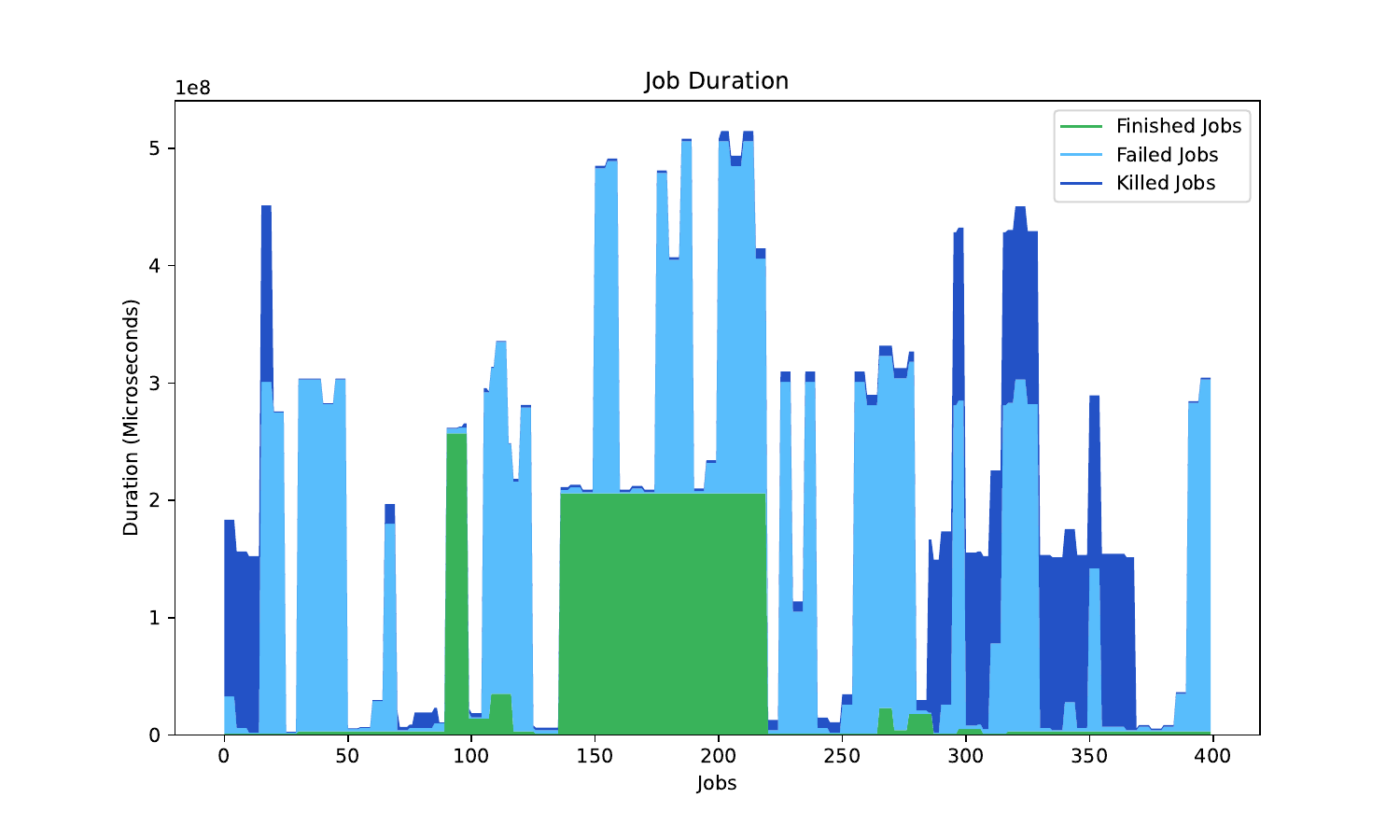}
  \caption{Job Duration Comparison}
  \label{fig:duration}
\end{figure}

As evident from the CPU usage distribution, the duration of a job is related to the end results (i.e., finished, failed, or killed). Failed jobs tend to run for longer periods and consume more resources. So, from our analysis, we see that around 40\% of the failed jobs run for a longer period than average jobs. The majority of the jobs finish within a fraction of the time. But the tasks that run for a longer period, fail in most cases. The same behavior was found for killed jobs as well. Few killed jobs had run longer than the failed jobs. Figure~\ref{fig:duration} shows the comparison of duration between failed, killed, and finished jobs.

\subsection{Job Resubmission}
\begin{figure}[]
  \centering
  \includegraphics[width=\linewidth]{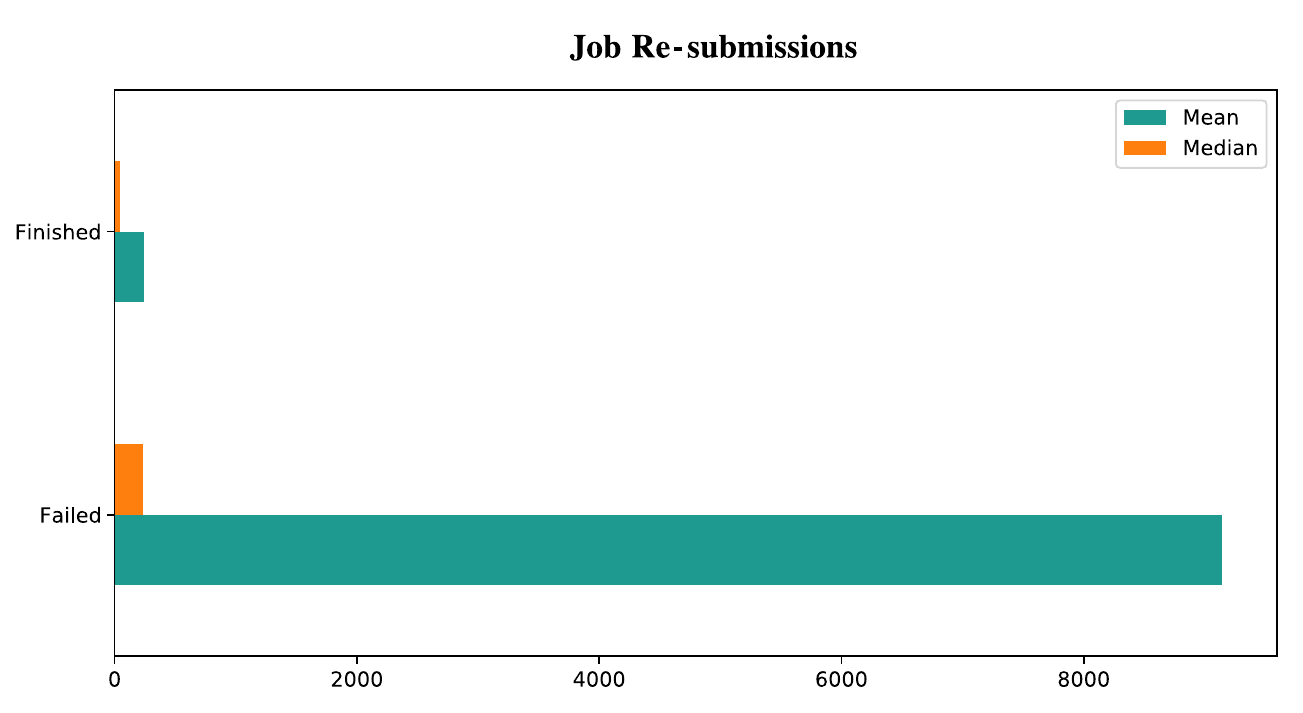}
  \caption{Job Re-submissions Comparison}
  \label{fig:re-submission}
\end{figure}

From our analysis, we found that job re-submission is a key factor in differentiating failed jobs. As evident from Figure~\ref{fig:re-submission}, failed jobs face much higher number of resubmissions than finished jobs. We found the highest resubmitted failed job was submitted 7947555 times. This is a great indicator of how much resources are wasted due to failed jobs.

\subsection{User Specific Failures}
\begin{figure}[]
\centering
  \hfill
  \subfigure[Jobs Submitted by Users]{\includegraphics[width=3.8cm]{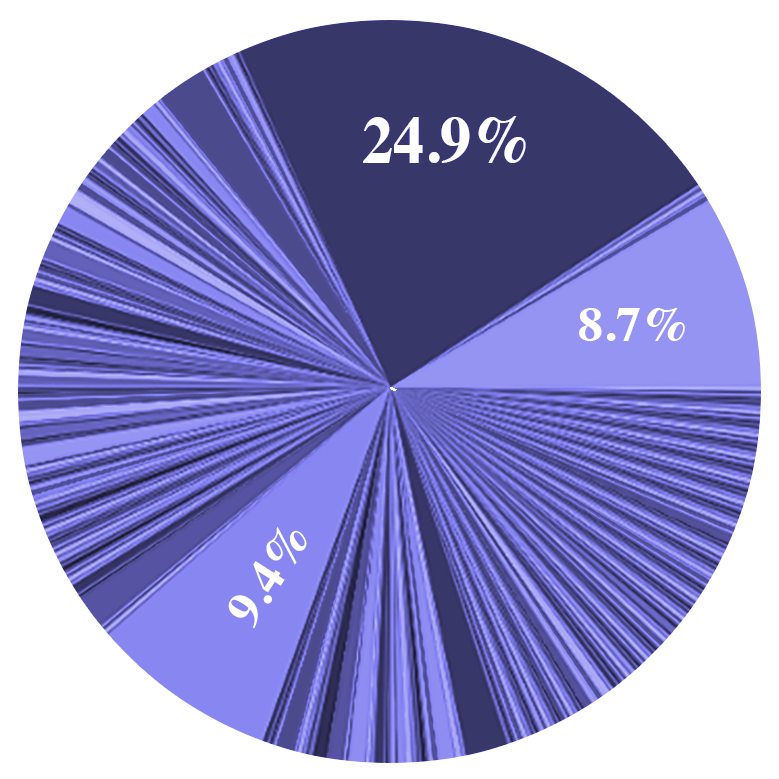}}
  \hfill
  \subfigure[Collections by Users]{\includegraphics[width=3.8cm]{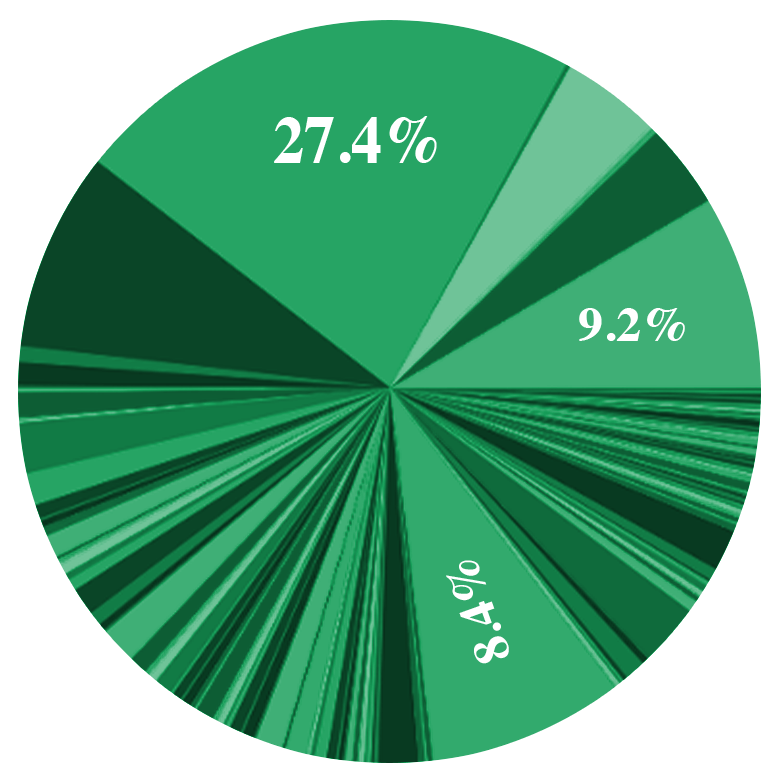}}
  \hfill
  \caption{Failed Jobs by Users}
  \label{fig:user-1}
\end{figure}

In this part, we tried to understand the user behavior and correlate them with resource consumption. In each cluster, all the submitted tasks are grouped into collections. There can be jobs and alloc sets in a collection. So, we tried to figure out the usual size of collections. As evident from Figure~\ref{fig:user-1}, 24.9\% of the failed jobs belong to a single collection. And there are some other large collections containing approximately 7-10\% of all failed jobs in a cluster. We imply that the number of re-submissions in these collections is higher. We also considered the job submission rate by individual users. We observed that approximately more than 20\% of the failed jobs in a cluster are submitted by a single user. And this is not an exception, there are some other users who have submitted a significantly higher percentage of total failed jobs. This gives us a clear understanding of the common usage patterns when there is a job failure.

\section{User Event Analysis}
Along with the behavior of collections, we are interested in knowing how users operate on the platform. We observe multiple cases where a user administered more than 50\% of the entire collection events on one cluster. We are able to track the top 5 users, who have dominantly consumed most of the resources.


\begin{itemize}
    \item \textbf{Top User 1:} This user controls 55.53\% of the overall collection events. Roughly 25\% of all events are KILL events in this case. This user is responsible for 75.50\% of all KILL events on the platform. All these jobs are production level jobs with relatively low latency sensitivities. In this case, auto-scaling option was used with user provided conditions. This is a potential weakness that needs to be addressed effectively while designing cloud schedulers. No single user should be allowed to dominate a platform that they can manipulate at will. 
    
    \item \textbf{Top User 2:} This user hypothesizes the effectiveness of using fully automated vertical scaling. User is running the same production level, low latency sensitive jobs that result in a close to 25\% FINISH rate and roughly 0\% KILL rate. 
    
    \item \textbf{Top User 3:} This user adopted best effort batch scheduler for all the jobs by using a mix of both constrained and fully automated vertical scaling. Also, we observed a similar variation in the scheduling class categories. 70\% of this user's jobs are latency sensitive and 30\% are not. 25\% of the events were QUEUE events, also indicative of user using a batch scheduler. FINISH and KILL percentages are similar to average rates for beb events. 
    
    \item \textbf{Top User 4:} This user validates the effectiveness of using fully automated vertical scaling for most of their jobs. There are also some events ($<$10\%) where this user has completely turned off vertical scaling. Most of their jobs are latency insensitive production level jobs that achieve a good FINISH rate of greater than 20\%. 
    
    \item \textbf{Top User 5:} This is the most interesting user amongst all five top users. The user has used fully automated scaling ($\sim$25\%), constrained scaling ($\sim$35\%), and completely turned off vertical scaling for 40\% of the events. More than 80\% of this user's events are high on latency sensitivity and almost all events are production level. As compared to other latency sensitive events, the KILL percentage is higher and FINISH rate is lower. 

\end{itemize}

This over-dominance of a few users possesses a threat to service reliability and availability as it could lead to queuing and scheduling delays. There should be a mechanism to prevent users from submitting mass collection events by either imposing a hard limit or promoting vertical scaling to prevent the under-utilization of resources. 

\section{Discussion and Future Directions}
In this paper, we performed a thorough analysis of the Google Cluster Trace 2019 dataset and found out several insights for failure characterization, resource usage, and user behavior. 
Our major research findings in this work are as follows: 

\begin{itemize}
  \item The average CPU usage of failed jobs is much higher than the finished jobs. Even in some cases, it is 13 times higher than finished jobs. 
  \item CPU usage becomes increasingly high with the percentile of total run-time for failed jobs. 
  \item On average, failed jobs use less memory than finished jobs. However, finished jobs have a diverse range of resource usage. 
  \item Approximately 40\% of the failed jobs run for a longer period of time than the average duration of the finished jobs.
  \item More than 20\% of the collections containing failed jobs are submitted by a single user in a cluster.
  \item There are multiple cases where a single user submitted more than 50\% of the entire collection events on one cluster. Vertical scaling could be used to make optimum use of available and requested resources. 
\end{itemize} 

The purpose of this research is to find out insightful characteristics of failed jobs. We believe these findings will be useful for developing a robust cloud scheduler or an early failure prediction model. Most of the failures can be handled properly with the combination of these three steps-- i) failure characterization, ii) early failure prediction, and iii) dynamic rescheduling based on the failure prediction results. Previously, several research has been conducted for developing failure prediction systems based on the Google cluster Trace 2011 and other datasets \cite{pavel2017}, \cite{dabbagh2015}, \cite{gao2019}. We think the characteristics that this research unveils will be crucial for developing a more advanced failure prediction model and optimized job scheduler.

\section{Conclusion}
This work analyzes the Google Cluster Workload Traces 2019 to uncover unorthodox findings by focusing on machine and collection-level events. We found some significant characteristics of failed jobs and also compared them with finished jobs to better understand the resource usage patterns in diverse cloud workloads. We also performed a novel user behavior analysis that unveils the heterogeneity in collection events and the over-dominance of a few users. We believe that the observations and analysis performed in this paper would be beneficial to understand the workload patterns in large-scale cloud data centers from multiple perspectives. It will also allow researchers to create more effective schedulers and failure prediction systems for ensuring cloud platform reliability and robustness. 


For future work, we wish to dig deeper into understanding the remarkable shift of free tier jobs to best-effort batch jobs that will help cater to the users’ requirement needs. We plan to employ parallel research to combine queuing delays for collections and scheduling delays for jobs to better visualize the job life cycle inside best effort batch tier jobs. Also, we plan to investigate the quantitative contributions of vertical scaling to reduce slack resources. 

\bibliographystyle{IEEEtran}
\bibliography{main}
\end{document}